\documentclass[a4paper,12pt,twoside]{article}
\usepackage{amsmath,amssymb,amsthm}
\usepackage{graphics,epsfig,calc}
\usepackage{upref}
\def\bx{\mathbf{x}}
\def\bq{\mathbf{q}}
\def\hx{{\hat{\mathbf x}}}
\def\hq{{\hat{\mathbf q}}}
\def\bk{\mathbf {k}}
\def\bl{\mathbf {l}}

\author{V.~S.~Buslaev$^{1}$, S.~B.~Levin$^{1}$, P.~Neittaanm\"aki$^{2}$, T.~Ojala$^{2}$}
\title{New approach to numerical computation of the eigenfunctions
of the continuous spectrum of three-particle Schr\"odinger
operator.\\ I. One-dimensional particles, short-range pair
potentials}
\date{}

\begin{document}

\maketitle

\begin{center}
{$^1$Department of Mathematical and Computational Physics,\\
St-Petersburg State University, Russia \\
$^2$Department of Mathematical Information Technology,\\ University
of Jyvaskyla, Finland}
\end{center}

\abstract{Basing on analogy between the three-body scattering
problem and the diffraction problem of the plane wave (for the case
of the short range pair potentials) by the system of six half
transparent screens, we presented a new approach to the few-body
scattering problem.
 The numerical results have been obtained for the case of
the short range nonnegative pair potentials. The presented method
allows a natural generalization to the case of the long range pair
potentials. }


\section{Introduction}

\subsection{}

The quantum system of two particles interacting via the Coulomb
potential is probably the most known model of the Quantum mechanics.
The model allows an explicit solution.  Oppositely, the mathematical
status of the system of three quantum particles with the pair
Coulomb interaction is relatively poor. The system of three
particles with short range pair interactions was successfully
studied by L. Faddeev \cite{Fad1}, but the direct generalization to
the Coulomb type potentials was found impossible.  Something,
however, is known: the quantative nature of the spectrum and the
asymptotic behavior of the solutions of the non-stationary
Schr\"odinger equation. These results were obtained in frameworks of
a non-stationary approach, see \cite{enss,derez}. Nevertheless  a
mathematically consistent stationary approach similar to the
Lippmann-Schwinger integral equation, or something analogous,  was
not developed though. Such an approach is needed if we are
interested in numerical parameters of many important physical
processes like dissociative recombination in atomic and molecular
physics with applications to astrophysics, formation and break up
processes of large molecules in bioengineering and medicine,
formation of the molecular resonance states in chemical physics,
dynamics of the few electron systems in wave-conductor
nano-technology.

There are specific difficulties that are characteristic for the systems with Coulomb type
interactions.

They are naturally explained by the fact that the long range
interactions crucially affects the asymptotic behavior at infinity
in the configuration space of the eigenfunctions, Green's functions
and other similar objects. The consequences of that affection on the
structure of asymptotics  up to now have not been taken in account
in correct mathematical manner. As a result, such approaches to many
particle scattering as Faddeev's equations \cite{Fad1}, AGS
equations \cite{ags67}, successfully applicable to the systems with
short range potentials, do not work for the systems with the long
range potentials.

The asymptotic behavior of the wave functions for the systems of few
charged particles has been studied only in some domains of
configuration space but not for all asymptotic directions. Let us
shortly list some known results. In \cite{p1,p2} there was studied
the asymptotic behavior of three charged particles wave function for
the case of large distances between all three particles. Another
limiting case, considered in \cite{p3}, corresponds to
configurations with one Jacobi coordinates been much larger than
another one.

In the list of the literature reflecting the theoretical aspects of
the problem we mention also \cite{Rudge}, \cite{Peterkop},
\cite{FaddeevMerkuriev}, \cite{Merkuriev} ,
\cite{am92,asz78,AltLevinYakovlev}. Application to computational
aspects of the problem were treated in \cite{as96,Oryu,Deltuva},
\cite{kvr01,bly,Suslov}, \cite{RescignoBray}.

One of the typical computational approaches to such systems is to
replace the Coulomb potentials by the Yukava potentials (or some
other cut-off potentials), to compute the parameters of the
scattering for this modified system, and to consider the results for
small screening parameter. Mathematically, it is not a completely
satisfactory procedure. Some other approximate approaches also
exist.

\subsection{}

We started a description of a new approach to the mathematically
consistent stationary treatment of the scattering in the systems of
three quantum particles with long range pair interactions in
\cite{BL}. We are going to consider in turn the case of three
one-dimensional particles with short range interaction (it is
already completed theoretically and  published), the case of three
one-dimensional particles with long range (Coulomb type)
interactions (it is also completed at the moment but is not
published yet) and the case of three three-dimensional particles.
Each next case will be based on the results for the preceding stage.
We hope that we will be able to illustrate each theoretical stage by
numerical computation of the field. This paper contains the
numerical results illustrating the formulas of paper \cite{BL}.

We assume here that the pair potentials are non-negative. In this
case the spectrum is purely continuous, covers the positive
semi-axis and is in the natural sense homogeneous. In fact, this
case is the most interesting at the present stage since the lower
spectral branches for negative total energy in case of charged
particles was already treated in \cite{Veselova}.

It is worth mentioning that the scattering in the system of
one-dimensional particles is not just a first step on the way to the
case of three-dimensional particles. It is interesting by itself,
the systems of three one-dimensional particles (neutral or charged)
were  intensively studied during many years (see, for example,
\cite{Yang,Lieb,McGuire,Olshanii,Green}). In recent years there
appeared a new interest to such systems since they were realized
experimentally (see \cite{Gorl,K1,K2,Esteve}).

The main idea is to suggest a priori explicit formulas for the asymptotic
behavior of the eigenfunctions of the continuous spectrum (for example, the
scattered plane waves).

The formulas describe the eigenfunctions at infinity up to the
simple diverging waves with  smooth amplitudes. If we are able to
find such asymptotic behavior (satisfying certain criteria that will
be discussed later on) even heuristically, we  obtain a way for
regular numerical computations of the eigenfunctions. We obtain
simultaneously also a method to construct an appropriate integral
equation of the same nature as the Lippmann-Schwinger equation for
the scattering of the plane wave by a quickly decreasing potential
that can be used  to justify the asymptotic behavior rigorously
following the ideas of \cite{BV}.

For one-dimensional particles with quickly decreasing at infinity
pair potentials we can use, for the description of the mentioned
asymptotic behavior, the analogy between the stated problem and the
classical problem of the diffraction of the plane waves by the set
of semi-transparent infinite screens. This analogy was already used
in \cite{BM,BMC,BMC1,BL}. In case of long range potentials we are
able to treat the diffraction problem analogously with the
replacement of the classical plane waves by plane waves that are
appropriately deformed by the long range tails of the Coulomb
potentials. It is important to mention that the diffraction itself
and the corresponding scattering problems cannot be completely
reduced to the scattering of the plane waves by the screens; we have
to add to these processes some genuine diffraction components that
have more complicated analytical structure  but still explicit
description. This more complicated  structure is also dictated by
the analogy with the classical diffraction theory.

Here we consider a system of three identical one-dimensional quantum
particles interacting via short-range pair potentials. These strict
limitations allow to  simplify  the narration and the view of the
formulas, but the essence of the main questions which we are
interested in and their treatments is not affected. In the following
parts we consequentially will get rid of these limitations. As we
have mentioned above, the theoretical part of this work is already
published, but we decided for the completeness to repeat shortly the
main theoretical ideas of \cite{BL}. The main goal of the work is to
confirm that the approach works for the numerical computation of the
eigenfunctions of the continuous spectrum. The approach is new even
for the short range pair potentials.

The structure of the work is as following: it consists of two parts.
The first part is devoted to the known theoretical constructions.
The second one is original and represents the results of numerical
computer computations.

\section{Main formulas}

\subsection{Configuration plane}

The configuration space of the system after the separation of motion
of the center of mass is the hyperplane  $\Gamma = \{\bx =
(x_1,x_2,x_3): x_1+x_2+x_3 = 0\}$ in $\mathbf{R}^3$. The
Schroedinger equation has the form:
\begin{equation}
-\triangle \psi + (v(x_1) + v(x_2) + v(x_3))\psi = E\psi,
\end{equation}
where $\psi = \psi(\bx) \in \mathbf{C}$, $\triangle$ is the Laplace
operator on $\Gamma$ that will be described  more specifically later
on. The  real-valued function $v(x),\ x \in \mathbf{R},$ is the
potential of the pair interaction. In the present text it is
supposed to be an even function with a compact support,  $v(x) = 0,
\ |x| > b/2.$ We suppose $E > 0$.

The scalar product on $\Gamma$  is given by the formula
\begin{equation}
<\bx,\bx'> = \frac{2}{3} (x_1x'_1 + x_2x'_2 + x_3x'_3).
\end{equation}
As usual, the norm of the vector is defined by the formula  $|\bx|^2 = <\bx,\bx>.$
The Laplacian is also generated by this scalar product.

Let us consider on  $\Gamma$ three straight lines $l_j = \{\bx: x_j
= 0\},\ j = 1,2,3,$ and three unit vectors  $\bl_j $ that belong to
these lines and oriented such that   $x_{j+1}$ increases  along
$\bl_j $. Consider also the unit vectors $\bk_j $ that are
orthogonal to  $\bl_j $ and oriented along the direction of
increasing of $x_j$. Consider, at last, three pairs of the cartesian
coordinates  $(x_j, y_j)$ with respect to the bases  $(\bk_j , \bl_j
)$. These are, so called, Jacobian coordinates on  $\Gamma$. With
these coordinates
\begin{equation}
<\bx, \bx'> = x_jx'_j + y_jy'_j, \quad |\bx|^2 = x_j^2 + y_j^2, \quad j=1,2,3,
\end{equation}
and
\begin{equation}
\triangle = \frac{\partial^2}{\partial x_j^2} + \frac{\partial^2}{\partial y_j^2}, \quad j=1,2,3.
\end{equation}

The lines  $l_j$ define on the plane $\Gamma$ six sectors. The
internal part of a certain one consists of the vectors
$(x_1,x_2,x_3)$ whose coordinates satisfy the condition
$x_{j_1}>x_{j_2}>x_{j_3}$ where $\sigma=(j_1,j_2,j_3)$ is a
permutation of the numbers  $(123)$. We will denote any sector by
the corresponding permutation  $\sigma$ and will write
$\lambda=\lambda_\sigma$ (see  Figure 1).

\vskip0.5cm
\hskip-3.5cm\includegraphics[scale=1.1,angle=0]{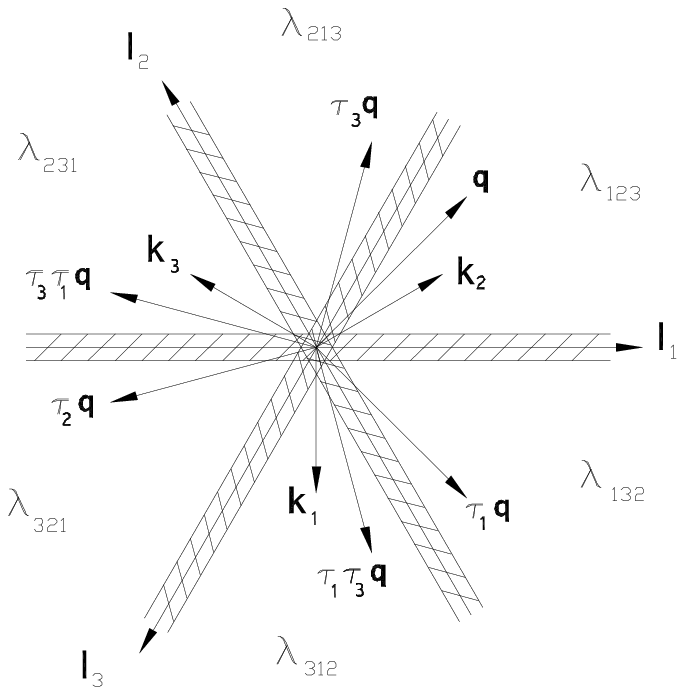}

\vskip0.3cm \hskip1cm  Figure 1

\vskip1.5cm

Let the group $S_3$ of  permutation acts on  $\Gamma$ so that
$$
(\sigma,\bx)\rightarrow \sigma\bx=(x_{j_1},x_{j_2},x_{j_3}),\ \ \ \
\sigma=(j_1,j_2,j_3),\ \ \ \ \bx=(x_1,x_2,x_3).
$$

The group contains 6 elements. The permutation can be identical, or
a transposition of two elements, or a composition of two
transpositions, some of the compositions coincide. Introduce the
notations for the transpositions: $\tau_1=(132)$, $\tau_2=(321)$,
$\tau_3=(213)$, and notice that $\tau_i^2=I,\ \ \ i=1,2,3$. The
action of the transposition on $\Gamma$ will be denoted by the same
symbol $\tau_j,\ \ j=1,2,3$. It corresponds to the reflection with
respect to the line   $l_j, \ \ j=1,2,3$. It is clear that
\begin{equation}
\tau_1 (x_1,x_2,x_3) = (-x_1,-x_3,-x_2), \quad \tau_1 (y_1,y_2,y_3)=
(y_1,y_3,y_2),
\end{equation}
the analogous formulas are also satisfied for  $\tau_2, \tau_3.$ The
composition of two transpositions generates a rotation, and the
following equalities, in particular, hold:
$\tau_1\tau_2=\tau_3\tau_1=\tau_2\tau_3$,
$\tau_2\tau_1=\tau_1\tau_3=\tau_3\tau_2$.

Six elements  $\tau$ of the group $S_3$ generate six vectors
$\tau\bq$. If $\bq\in\lambda_\sigma$ then $\tau\bq\in
\lambda_{\tau\sigma}$.


\subsection{Separation of variables}

Consider now the eigenfunction that describes the scattering in the
system where just one of three potentials is not equal to zero.  Now
we deal with the Schr\"odinger equation
\begin{equation}\label{ODE}
-\triangle \chi_j + v(x_j)\chi_j = E\chi_j.
\end{equation}
It allows the separation of variables:
\begin{equation}
\chi_j (\bx, \bq) = \chi(x_j, k_j)e^{ip_jy_j} .
\end{equation}
The sense of the variables  $(x_j,y_j)$ is clear, $(k_j,p_j) $ are
the Jacobian coordinates of a given  vector  $\bq$.

The function $\chi(x, k), x,k \in \mathbf{R},$ is a solution of the ordinary differential equation
\begin{equation}\label{ode}
-\chi_{xx} + v(x)\chi = k^2 \chi,
\end{equation}
it has to be described separately. For $k > 0$ there exists and is
unique the solution that is characterized by the following
asymptotic behavior:
\begin{equation}
\chi(x,k)\sim s(k)e^{ikx},\  x \to +\infty; \quad \chi(x,k)\sim
e^{ikx} + r(k)e^{-ikx},\  x \to -\infty.
\end{equation}
On the whole axis $k$ this solution, due to the evenness of the potential, has to be
extended by the formula  $\chi(x,k) = \chi(-x,-k)$. Here  $s$ and $r$ are some complex-valued
functions of $k$ that are called the transition and the reflection coefficients.

We will suppose here that $v(x) \geq 0$ therefore the equation (\ref{ode}) does
not have the  bound states.


\subsection{Formal setting of the problem}

Our final goal is to construct the solution  $\psi(\bx,\bq)$ of the
Schr\"odinger equation that is characterized by the following
behavior at infinity:
\begin{equation}
\psi = n(\hx,\bq) \frac{e^{-i|\bx||\bq}|}{|\bx|^{1/2}} +
f(\hx,\bq) \frac{e^{i|\bx||\bq|}}{|\bx|^{1/2}} + o\left(\frac{1}{|\bx|^{1/2}}
\right), \quad \hx = \frac{\bx}{|\bx|}.
\end{equation}
Here
\begin{equation}
n(\hx,\bq) = \sqrt{\frac{2\pi}{i|\bq|}}\delta(\hx, \hq),
\end{equation}
and the $\delta$ function has to be considered with respect to the angle measure on the unit circle.

The asymptotic behavior has to be treated in a weak sense (in sense of distributions) with respect to $\hx$.
The coefficient $n$ before the converging circle wave coincides with the analogous coefficient before the
converging wave in the weak asymptotic representation of the plane wave
$e^{i<\bx,\bq>}$. Therefore the solution $\psi(\bx,\bq)$ can be naturally called the scattered plane wave.

Due to the symmetries of the potential $\psi(\bx,\bq) =
\psi(\sigma\bx,\sigma\bq),\ \sigma \in S,$ we always can assume that
$\bq$ belongs to a certain sector, say
$\lambda_I\equiv\lambda_{123}$. We restrict ourselves here by the
assumption that $\bq$ does not belong to neighborhoods of the
boundaries of the sector. It would be not hard to consider also the
case when $\bq$ belongs to the lines $l_j$ and their neighborhoods.

The function $f$ is a singular distribution. We will see that it has
singularities on all six directions $\sigma \bq,\ \sigma \in S$.
Four of them are of $\delta$ - function type, two (for $\sigma\bq =
\tau_2\tau_3\bq,\  \tau_2\tau_1\bq$) are of type of Cauchy's
limiting kernel. It is worth to notice that although the asymptotic
behavior is singular the solution itself is, naturally, a smooth
function.

In the case of the scattering by a quickly decreasing at infinity potential the asymptotic
behavior is given by the formula
\begin{equation}
\psi(\bx,\bq) = e^{i<\bx,\bq>}
+ f(\hx, \bq)\frac{e^{i|\bx||\bq|}}{|\bx|^{1/2}}
+ o\left(\frac{1}{|\bx|^{1/2}}
\right),
\end{equation}
where that time the scattering amplitude $f$ is not a singular distribution,
but a smooth function, and the asymptotic behavior can be treated in uniform sense.

Under our assumptions over the potential the scattered plane waves
create for  $E>0$ a complete system of the eigenfunctions of the
uniform in multiplicity continuous spectrum of the three particle
Schr\"odinger operator, $E \geq 0$.

Our further plan is following: we construct in explicit form a function  $\psi_1(\bx,\bq)$
and hope that the difference   $\psi - \psi_1$  has the diverging  asymptotic behavior
\begin{equation}\label{smooth_g}
\psi(\bx,\bq) - \psi_1(\bx,\bq) =  g(\hx,\bq) \frac{e^{i|\bx||\bq|}}{|\bx|^{1/2}} +
o\left(\frac{1}{|\bx|^{1/2}}
\right),
\end{equation}
where $g$ is a continuous function of the arguments.

Constructing  $\psi_1$ we use two criteria:
(1) The discrepancy
\begin{equation}
Q[\psi_1](\bx,\bq) = -\triangle \psi_1 + (v(x_1)+ v(x_2)+ v(x_3))\psi_1 - E\psi_1, \quad E
 = |\bq|^2.
\end{equation}
sufficiently quickly vanishes at infinity, (2) The asymptotic
representation for $\psi_1 - e^{i<\bx,\bq>}$ contains asymptotically
only the diverging wave.

Consider the difference
\begin{equation}
\xi = \psi - \psi_1.
\end{equation}
It satisfies the equation
\begin{equation}\label{eq}
H\xi - E\xi = -Q, \quad H = -\triangle  + (v(x_1)+ v(x_2)+ v(x_3)).
\end{equation}
Since $Q$ is quickly vanishing one can hope that $\xi$ asymptotically behaves as the diverging wave
\begin{equation}\label{rc}
\xi(\bx, \bq) = g(\widehat{\bx}, \bq)
\frac{e^{i|\bq||\bx|}}{|\bx|^{1/2}} +
o\left(\frac{1}{|\bx|^{1/2}}\right),
\end{equation}
with a continuous amplitude $g$. In other words, $\xi$ satisfies the classical radiation conditions at
infinity.

Further, it is naturally to hope that for $\xi$ we can construct an
integral equation with the same properties as the properties of
classical Lippmann-Schwinger equation. We can do it developing the
ideas of work \cite{BV}. However, preliminary, we can try to use
(\ref{eq})-(\ref{rc})  for the numerical computation of $\xi$ and,
consequently, of $\psi$. For the numerical computations we can
replace (\ref{rc}) by approximate boundary condition
\begin{equation}\label{rc1}
\left(\frac{\partial}{\partial |x|} - i\sqrt{E}\right)\xi =0\,,\ \
\text{for}\ \  |x| = R,
\end{equation}
where $R$ is sufficiently large.

The following construction of $\psi_1$ will consist of two steps. At
first, we construct for $\psi_1$ so called ray approximation
$\psi_R$. Its discrepancy has some singularities. After a natural
modification motivated by some classical diffraction problems the
discrepancy will become a smooth function.


\subsection{Ray approximation}

Consider six vectors $\sigma\bq$. These vectors, more precisely,
spanned by them rays, separate six sectors that we denote
$K_j^{\pm}$. The indices of the notation coincide with the indices
of the vector $\pm \bl_j$ that belongs to the sector $K_j^{\pm}$.

Now we can give  explicit expressions for the ray approximation in different sectors  $K_j^{\pm}$.

\underline{Sector  $K_1^+$:}  $\psi_R=\psi_1^+$,
$$
\psi_1^+(\bx,\bq) = \chi_1(\bx, \bq)s_2s_3.
$$
We use here the following notations: $s_j=s(k_j),\ \ r_j=r(k_j)$.

\underline{Sector $K_3^-$:}  $\psi_R=\psi_3^-$,
$$
\psi_3^-(\bx,\bq)=\chi_3(\bx,\bq)s_1s_2.
$$

\underline{Sector  $K_2^+$:}  $\psi_R =\psi_2^+$,
$$
\psi_2^+(\bx,\bq)=\chi_2(\bx,\bq)s_1+ \chi_2(\bx,\tau_3\bq)s_2r_3.
$$

\underline{Sector $K_2^-$:}  $\psi_R=\psi_2^-$,
$$
\psi_2^-(\bx,\bq)=\chi_2(\bx,\bq)s_3+ \chi_2(\bx,\tau_1\bq)s_2r_1.
$$

\underline{Sector $K_1^-$:}  $\psi_R =\psi_1^-$,
$$
\psi_1^-(\bx,\bq)=\chi_1(\bx,\bq)+ \chi_1(\bx,\tau_2\bq)r_2s_1
+\chi_1(\bx,\tau_3\tau_1\bq)r_2r_1+ \chi_1(\bx,\tau_3\bq)r_3.
$$

\underline{Sector $K_3^+$:}  $\psi_R =\psi_3^+$,
$$
\psi_3^+(\bx,\bq)=\chi_3(\bx,\bq)+ \chi_3(\bx,\tau_2\bq)r_2s_3
+\chi_3(\bx,\tau_1\tau_3\bq)r_2r_3+ \chi_3(\bx,\tau_1\bq)r_1.
$$

The total field $\psi_R$ is defined by the formula
$$
\psi_R=\theta_1^+\psi_1^++\theta_3^-\psi_3^-+ \theta_2^+\psi_2^+
+\theta_2^-\psi_2^-+ \theta_1^-\psi_1^-+\theta_3^+\psi_3^+.
$$

The notation $\theta_j^{(\pm)}$ is used here for the characteristic function of the
corresponding sector $K_j^\pm$,
$$
\theta_1^++\theta_3^-+\theta_2^++\theta_2^-+\theta_1^-+\theta_3^+=1.
$$

In this formula the value of the field $\psi_R$ on the boundaries of
the sectors is not defined. In  \cite{BL} it was shown that on all
boundary rays except two, directed along the vectors
$$
\bq_{23} \equiv\tau_2 \tau_3 \bq,\ \ \ \bq_{21}\equiv \tau_2 \tau_1
\bq,
$$ the field is smooth, and its discrepancy everywhere except the
two vectors is equal to zero.


\subsection{Diffraction corrections}

The diffraction corrections on rays directed along the vectors
$\bq_{23}$ and $\bq_{21}$, can be constructed quite easily. Consider
the sector $\lambda_{231}$ containing $\bq_{23}$. Introduce the
polar coordinates $(r= |\bx|, \omega)$. Let us orient the angle from
$\bl_2^+$ to $\bl_1^-$. Let $\omega_{23}$ correspond to $\bq_{23}$.
Introduce four angles $0 < \omega_1 < \omega_2 < \omega_{23} <
\omega_3 < \omega_4 < \pi/3$. Consider the open covering of the
interval  $(0, \pi/3)$ by the subintervals $(0, \omega_2)$,
$(\omega_1, \omega_4$), $(\omega_3, \pi/3)$ and introduce a
subordinated partition of unit:
\begin{equation}
1 = \zeta_1 + \zeta_2 + \zeta_3.
\end{equation}

Further consider the function
\begin{equation}
\Phi (\alpha) = \frac{e^{-i\frac{\pi}{4}}}{\sqrt{\pi}} \int_{\infty}^{\alpha}e^{it^2}dt.
\end{equation}
Notice that
\begin{equation}
\Phi (\alpha) \to 1, {\text {as}}\,\, \alpha \to +\infty , \quad
\Phi (\alpha) \to 0,\ {\text {as}}\,\, \alpha \to -\infty.
\end{equation}

In more detail:
$$ \Phi (\alpha) =
1 + \frac{e^{-i\frac{\pi}{4}}}{\sqrt{\pi}} \frac{e^{i\alpha^2}}{2i\alpha} + \Delta\Phi(\alpha), \quad
 \Delta \Phi(\alpha) = - \frac{e^{-i\frac{\pi}{4}}}{\sqrt{\pi}} \int_{\alpha}^{\infty}\frac{e^{it^2}}
 {2it^2}dt = O(\alpha^{-3}),\ \ {\text {when}}\,\, \alpha \to +\infty.$$

Introduce the function
\begin{equation}
 \Phi^{(23)}_1 = \Phi(sign(\omega_{23} - \omega)||\bq_{23}||\bx| - <\bq_{23},\bx>|^{1/2}),
\end{equation}
\begin{equation}
\Phi^{(23)}_2 =  \Phi(sign(\omega - \omega_{23})||\bq_{23}||\bx| -
<\bq_{23},\bx>|^{1/2}) .
\end{equation}
It is known that
\begin{equation}
\phi = e^{i<\bx,\bq_{23}>}\Phi^{(23)}_j
\end{equation}
satisfies the Helmholtz equation  $-\triangle \phi - E\phi = 0.$

Now we can describe the diffraction corrections to the ray
approximation on $\lambda_{231}$. For that the ray field  $\psi_R =
\theta_2^+ \psi_2^+ + \theta_1^- \psi_1^-$ in the sector
$\lambda_{231}$ is replaced by
\begin{equation}
\psi_D^{(23)} = \psi_R    + \zeta_2 e^{i<{\bq}_{23},\bx>} [R_1  (\Phi^{(23)}_1 - \theta_2^+) +
R_2 (\Phi^{(23)}_2 - \theta_1^-)].
\end{equation}
$R_1 = r_1s_2r_3, R_2 = r_3r_2s_1 + s_3r_2r_1$.

Notice that the field  $\psi_R$ on the interval  $(\omega_1, \omega_4)$ contains the
discontinuous component
\begin{equation}
\psi_J = e^{i<{\bq}_{23},\bx>}[\theta_2^+ R_1  + \theta_1^- R_2],
\end{equation}
so the sense of the modification in nothing else but a simple
replacement of this discontinuous on $\bq_{23}$ component by a
smooth solution of the Helmholtz equation that outside of
$(\omega_2,\omega_3)$ gradually transfers to the original
discontinuous component up to a diverging circle wave with a smooth
amplitude. Outside of the interval  $(\omega_1,\omega_4)$ the
function $\psi_D$ coincides with the original ray approximation
$\psi_R$.

Analogous constructions can be also considered in the sector
$\lambda_{312}$. It is also worth to introduce here the polar
coordinates, and again to suppose that the angle $\omega$ varies in
the same limits with the same orientation, from  $\bl_3$ to
$-\bl_2$. We again can introduce the angles  $\omega_{21},\
\omega_j,\  j=1,2,3,4$ and a cutoff function $\zeta_2.$ After that
the modified field on $\lambda_{312}$ can be described by the
formula
\begin{equation}
\psi_D^{(21)} = \psi_R    + \zeta_2 e^{i<{\bq}_{21},\bx>} [R_2
(\Phi^{(21)}_1 - \theta_3^+) + R_1 (\Phi^{(21)}_2 - \theta_2^-)].
\end{equation}
Here
\begin{equation}
 \Phi^{(21)}_1 = \Phi(sign(\omega_{21} - \omega)||\bq_{21}||\bx| - <\bq_{21},\bx>|^{1/2}),
\end{equation}
\begin{equation}
\Phi^{(21)}_2 =  \Phi(sign(\omega - \omega_{21})||\bq_{21}||\bx| -
<\bq_{21},\bx>|^{1/2}) .
\end{equation}

As a result everywhere on $\Gamma$ outside of some circle $C_{r_1}$
with the center at $0$ and the radius $r_1$ there appears  a smooth
approximate wave field $\psi_0$:
\begin{equation}
\psi_0 = \psi_R \theta_I  + \psi_D^{(23)}\theta_{231}  +
\psi_D^{(21)}\theta_{312}.
\end{equation}
Here $\theta_{231}$ and $\theta_{312}$ are the characteristic
functions of the corresponding $\lambda$-sectors, and $\theta_I$ is
the characteristic function of their complement. Again there are no
jumps on the boundaries of the $\lambda$-sectors.

Consider a circle with the center at the origin. The radius $r_1$ of
this circle is defined by the condition that outside of the circle
on the rays directed along the vectors $\sigma\bq$ the sum of the
pair potentials is equal to zero.  Under this condition the field
$\psi_0$ can be additionally modified with the help of the cutoff
function  $\zeta(|\bx|)$ that is equal to $0$ for $|\bx| < r_1$ and
to $1$ for $|\bx| > r_2$ where $r_1 < r_2$.

The final expression for the approximate field  is now
\begin{equation}
\psi_1 = \psi_0\zeta.
\end{equation}


\subsection{Discrepancy}

We remember that there were proposed two criteria that have to be
taken into account when constructing the function  $\psi_1$. It is
sufficiently clear that the second one : (2) The difference $\psi_1
- e^{i<\bx,\bq>}$ contains asymptotically (in the weak sense) only
the diverging circle wave, is fulfilled. It is remained to check the
first one: (1) The discrepancy
\begin{equation}
Q[\psi_1](\bx,\bq) = -\triangle \psi_1 + (v(x_1)+ v(x_2)+ v(x_3))\psi_1 - E\psi_1, \quad E
 = |\bq|^2.
\end{equation}
sufficiently quickly vanishes at infinity.

From the previous formulas it follows that outside a certain circle of the radius $r_1$ the
discrepancy is not equal to zero only on some neighborhoods the rays generated by the vectors
$\bq_{23}$ and $\bq_{21}$. On these neighborhoods the discrepancy vanishes as $|\bx|^{-5/2}$.
It follows from this that the relative scattering amplitude $g(\hx,\bq)$, see (\ref{smooth_g}),
 must be continuous. Here we  give for the discrepancy a formula that can be used for the numerical
 computations of  $\psi$.

Consider now the field  $\psi_1$ on the neighborhoods of $\bq_{23}$
and $\bq_{21}$. It is not hard to see that the discrepancy of this
expression is equal to zero on the sectors where there is equal to
zero the derivative of the function  $\zeta_2$. It means that the
discrepancy $Q[\psi_0]$ can differ from zero only on the
subintervals  $(\omega_1, \omega_2)$ and $(\omega_3, \omega_4)$.
That implies that the discrepancy $Q^{(23)}$ on the sector
$\lambda_{231}$ can be naturally represented as the sum:
\begin{equation}
Q^{(23)}   = Q_1^{(23)} + Q_2^{(23)}.
\end{equation}
Similarly, on the sector $\lambda_{312}$
\begin{equation}
Q^{(21)}   = Q_1^{(21)} + Q_2^{(21)}.
\end{equation}

All four terms here can be easily computed. The answers are completely analogous. In particular,
\begin{equation}
Q_1^{(23)}  = R_1 (-\Delta - E) e^{i<\bq_{23},\bx>} (\Phi^{(23)}_1 -1)\zeta_2' =
\end{equation}
\begin{equation}
= R_1 [e^{i<\bq_{23},\bx>} (\Phi^{(23)}_1 -1) \frac{-1}{r^2}\zeta_2'' -
2i \frac{1}{r}\zeta_2 '<\bq_{23}, w> e^{i<\bq_{23},\bx>} \Delta\Phi^{(23)}_1] .
\end{equation}
where $w$ is a unit vector orthogonal to $\hx$ and
oriented along the direction of increasing $\omega$.

Finally,
\begin{equation}
Q[\psi_0] = Q_1^{(23)} + Q_2^{(23)} + Q_1^{(21)} + Q_2^{(21)}.
\end{equation}

It is easy to see that all four components of the discrepancy vanish at infinity like  $|\bx|^{-5/2}.$\\

The previous computations of the discrepancy were given for not
small $|\bx|$ where the supports of three potentials are separated.
Let us modify now the field $\psi_1$ by introducing in it the factor
$\zeta = \zeta(|\bx|)$ that is equal to $0$ for $|\bx| < r_1$, and
is equal to $1$ for $|\bx| > r_2$, $0 < r_1 < r_2.$ It is supposed
that for $|\bx| > r_1$ three supports do not intersect. The
definition of $\psi_1$ is given by the formula
\begin{equation}
\psi_1 = \psi_0 \zeta.
\end{equation}

The final expression for the discrepancy  is given by the formula
\begin{equation}
Q[\psi_1] = Q[\psi_0]\zeta - 2\left[\frac{\partial}{\partial
|\bx|}\psi_0(\bx,\bq)\right]\zeta^{'} - \psi_0
\frac{1}{|\bx|}\frac{\partial}{\partial |\bx|}|\bx|
\frac{\partial}{\partial |\bx|}\zeta.
\end{equation}
There is no problem in explicit computation of the derivative
$\frac{\partial}{\partial |\bx|}\psi_0(\bx,\bq)$.


\section{Numerical computations}


The goal of the computations was to show that the suggested plan is realistic and can be practically
used for the computations of the scattered plane wave and the corresponding amplitude of scattering.
The pair-particle potential $v(x)$ and the vector $\bq$ are two parameters of the problem.

As for $v(x)$
we choose the potential function
\begin{equation}
v(x)=\left\{
  \begin{array}[]{lc}
    2 e^{\frac{1}{(4 x)^2-1}+1},& |x|<\frac{1}{4} \\
    0 ,& \text{otherwise}.
  \end{array}
  \right.
  \label{equ:potential}
\end{equation}
Any specific choice is not crucial, we could take arbitrary even
potential (even non-necessary continuous) with the compact support.
With this potential we computed the solution $\chi(x,k)$ of
one-dimensional Schr\"odinger equation (\ref{ODE}) and found the
corresponding transition $s(k)$ and reflection $r(k)$ coefficients.

Then the solutions $\chi(x,k)$ were interpolated to the actual
computational domain to construct the functions $\chi_j$. This interpolation was
necessary only on the support of the potentials. Outside
of the supports the analytic expressions of the functions $\chi(x,k)$ were known
after the coefficients $s(k)$ and $r(k)$ were found
numerically.

We took $E = 4$. For the vector $\bq$  we used two choices: 1) $k_1
= 1,\ \ p_1 = \sqrt{3},\ \ $ 2) $k_1 = p_1 = \sqrt{2}.$ In the first
case the field as a function of $\bx$ is symmetric with respect to
the straight line generated by $\bq$. It was taken for the control.

The function $\psi_R$ was computed directly with the knowledge of
$\chi_j$, the Fresnel integral was taken from GSL (Gnu scientific
library).

The functions $\psi_0(\bx,\bq)$ and $\psi_1(\bx,\bq)$ were computed
with the help of the explicit formulas for them. The discrepancy $Q$
was also computed with the help of the explicit formulas. The radii
$r_1 < r_2$ were taken as $r_1 = 4,\ \ r_2 = 14.5.$ We think that
this choice reasonably corresponds to the selected value of $|\bq|$.

For the diffraction corrections (and near the origin) the chosen partition of unity
corresponds to function
$\zeta(z)=z^3(10 -15 z + 6 z^2), \; 0<z<1$, where $z$ is the variable relative to angle $\omega$.

Then we finally considered the boundary problem (\ref{eq} -
\ref{rc1}). Of course, it was the main part of the numerical program
of the work. The problem on the disc is not on the spectrum.

For the computations we used mainly FreeFem++, which is a user friendly language
dedicated for solving partial differential equations
with the finite element method. All the necessary steps from mesh creation to
solving the linear system can be done
within the same program in a manner that is not of a black box type. Since we
used the finite element method,
we introduced
the corresponding weak formulation of the problem:
find $\xi \in H^1(\Omega)$ such that
\begin{multline}
  \int_\Omega \nabla \xi \cdot \nabla w + (v(x_1)+v(x_2)+v(x_3)-E)\xi w \; dx
    - \int_{\partial \Omega} i \sqrt{E} \xi w \; dS  \\
    = - \int_\Omega Q w \; dx
      \quad \forall w \in H^1(\Omega)
      \label{equ:weak}
\end{multline}

The finite element discretization of (\ref{equ:weak}) was then done
in a standard fashion using quadratic Lagrange elements on a
triangular mesh. The computational domain was divided into
sub-domains to have the finite element mesh fit better with the
support of the potential $V$ and the constructed function $\chi_0$
and the discrepancy $Q$. A relatively uniform mesh was introduced
with lengths of triangle edges between 0.15 and 0.48. With a
circular domain of radius 190, the total number of degrees of
freedom was 3 million. We used Matlab's solver for large linear
systems.

The results are represented by the Fig.2-3,  and we think that they
are reasonable.

\vspace{5mm}

\vskip1.5cm \hskip3.5cm\includegraphics[scale=0.7,angle=0]{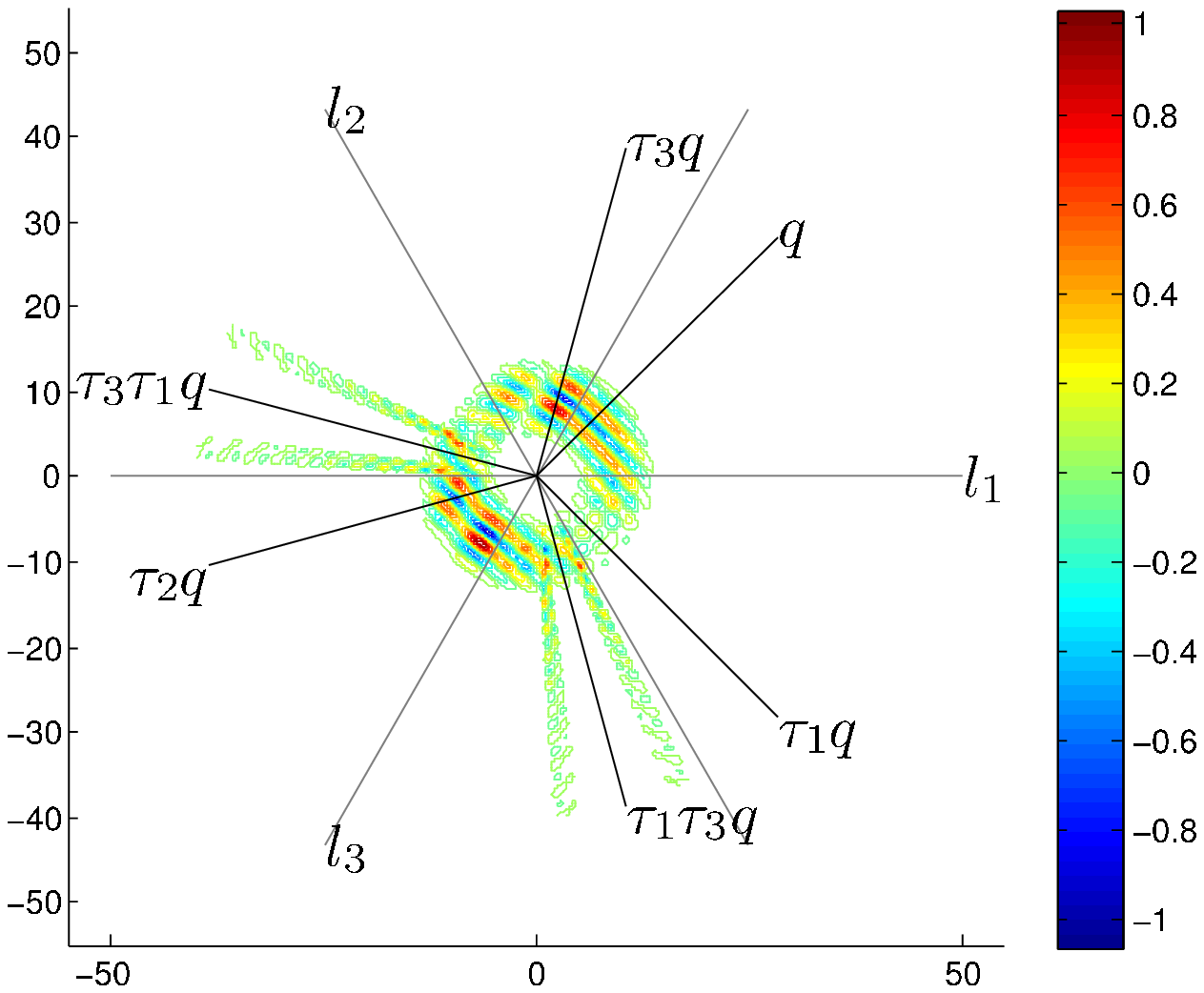}

\vskip-4cm \hskip1cm  Figure 2

\vskip4cm \hskip7cm $real(Q)$

\vskip4cm

\vspace{5mm}

\vskip-2.5cm \hskip3.5cm\includegraphics[scale=0.7,angle=0]{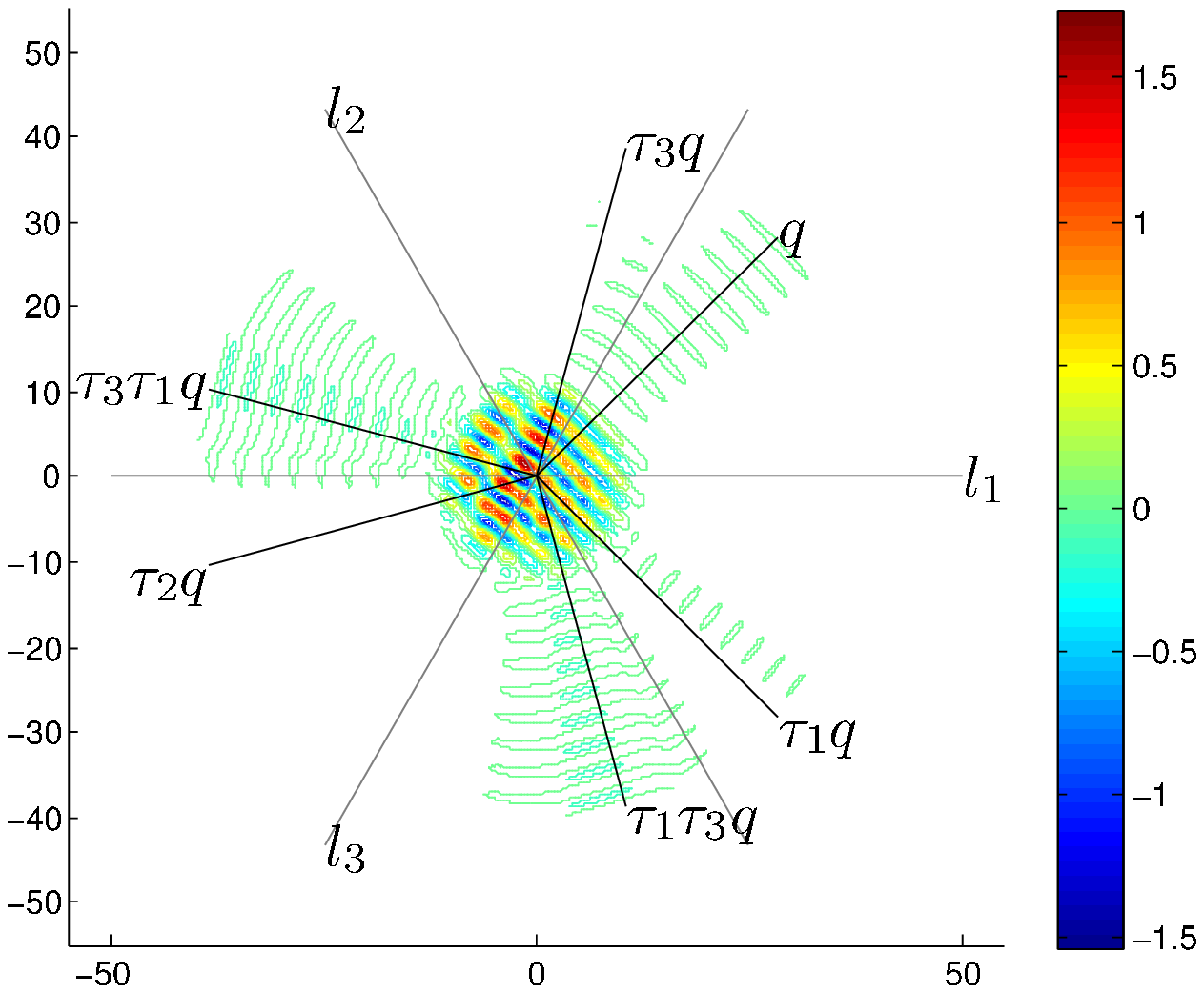}

\vskip-4cm \hskip1cm  Figure 3

\vskip4cm \hskip7cm $real(\xi)$





\vskip1cm

A certain problem was the choice of the radius $R$. To get more precise results it would be better
to take bigger $R$, but the bigger $R$ means the harder computations. The criterium of the
compromise was connected with the integral form of the radiation conditions. These conditions  are:\\
1)the integral
\begin{equation}
\int_{S_R}ds |\xi(\bx,\bq)|^2
\end{equation}
over the circle $|\bx|= R$ must be bounded for large $R$;\\
2) the integral
\begin{equation}
\int_{S_R}ds |(\frac{\partial}{\partial |\bx|} -
i\sqrt{E})\xi(\bx,\bq)|^2
\end{equation}
must decrease as $R^{-2}$.

Notice that Fig. 4 shows that the first integral here is
asymptotically approaching a constant at sufficiently large $|\bx|$,
and the second integral is quite small for such $|\bx|$, but does
not decrease for the present computations with $R=190$.

\vspace{5mm}

\vskip0.5cm
\hskip-0.5cm\includegraphics[scale=1.0,angle=0]{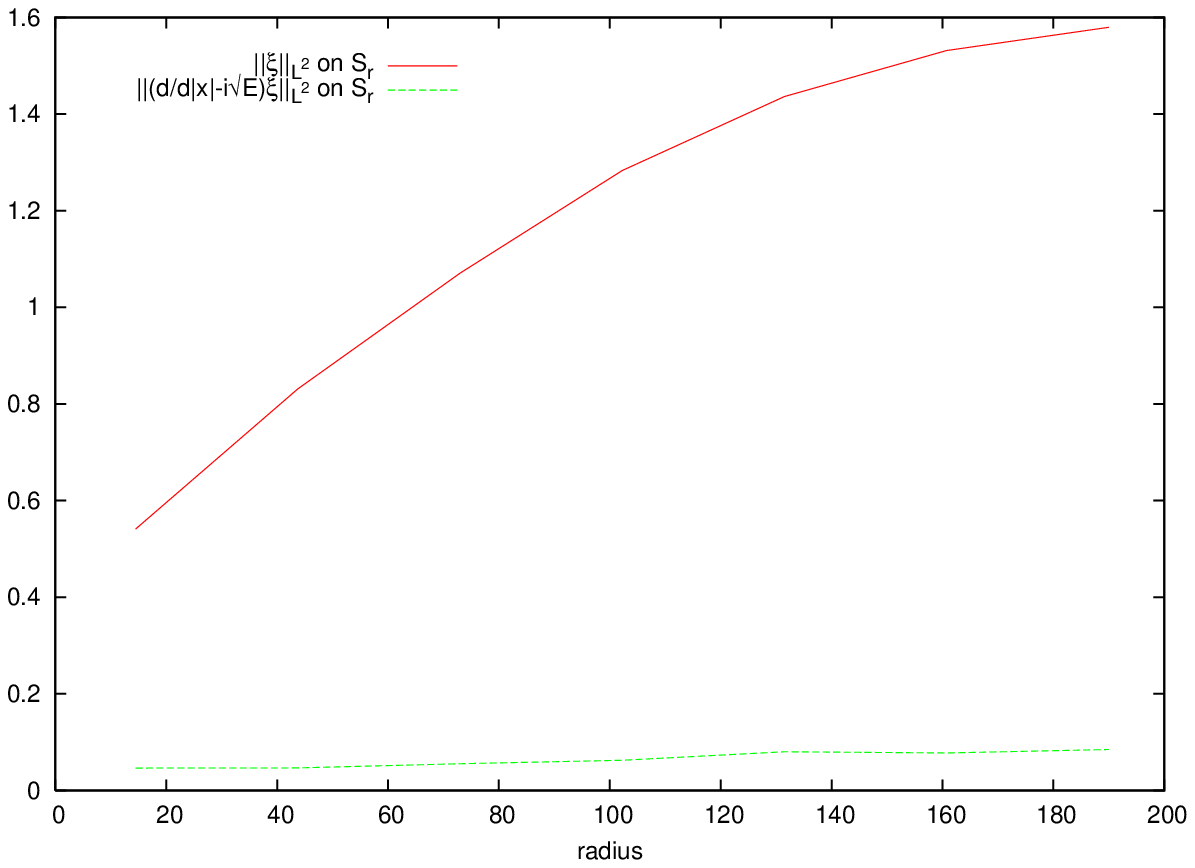}

\vskip1cm \hskip1cm  Figure 4

\vskip2cm

It, probably, means that such radius is not completely sufficient
for the final computations. However, the bigger radius would mean the harder
computations, so we decided at the moment to restrict the  radius of
the circle by $190$.

We considered also the corrected boundary condition where the next
term of asymptotic behavior of $\xi$  was also taken into account:
$$\left(\frac{\partial}{\partial r} - i|\bq| + \frac{1}{2r}  \right)\xi|_{r = R} = 0. $$
Nevertheless, the correction did not help to stabilize the
calculation in smaller domain, as it could be expected. The reason is that the term $\frac{1}{2R}$
appeared to be very small comparatively with other terms.

To clarify further the situation with the stabilization of $L_2$ - norm on the boundary
of the disk we should come back to behavior of
$\xi$ as a function of the angle at fixed radius $r= R$.

\vspace{5mm}

\vskip0.5cm
\hskip-0.5cm\includegraphics[scale=0.9,angle=0]{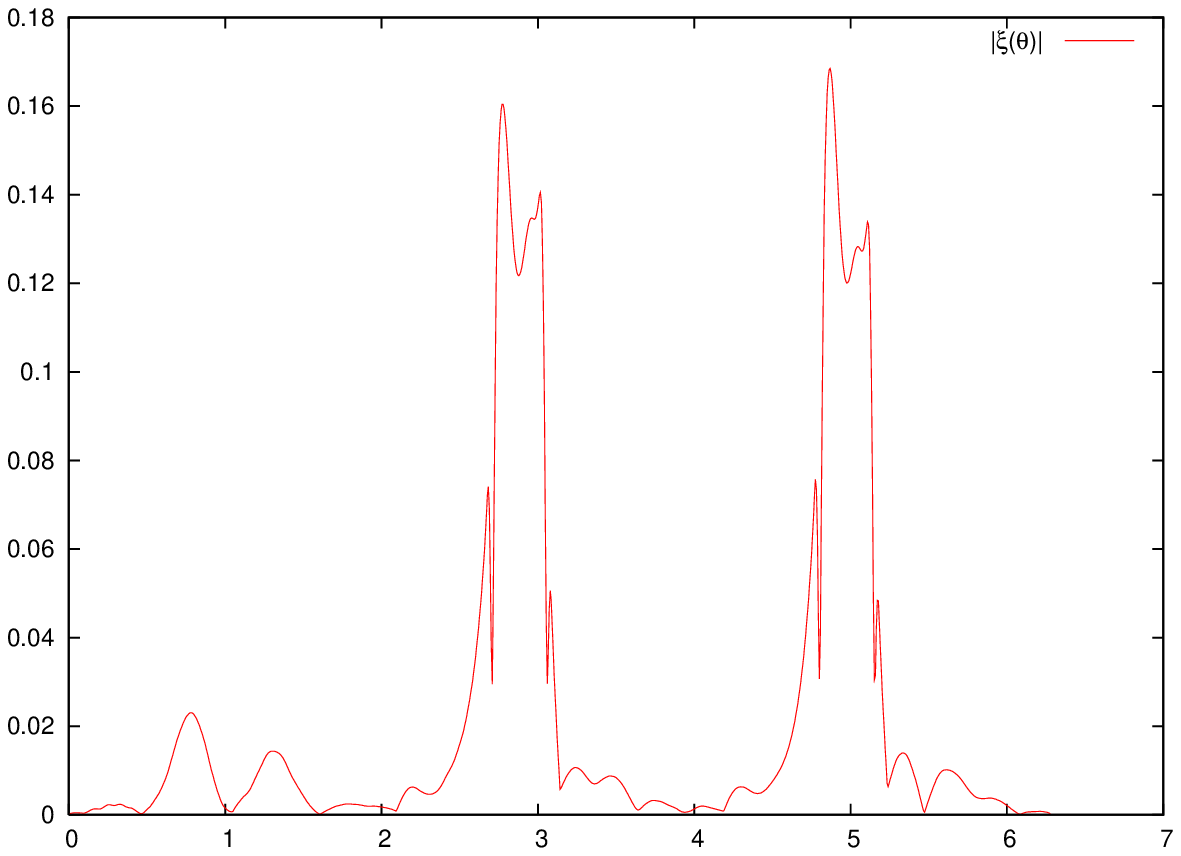}

\vskip1cm \hskip1cm  Figure 5 \hskip4cm $|\xi(\theta)|$

\vskip2cm

One can easily see from the Figure that the main contribution to the
integral comes from two special directions on the boundaries of
shadow and light in sectors $\lambda_{231}$ and $\lambda_{312}$, see
Fig.5.  Therefore for the stabilization of the whole integrals first
of all the contributions to them of two indicated sectors must be
stabilized.  Namely their stabilization was not completely reached
for considered size of configuration domain and requires bigger
scale of radii. There are also other computational indications that
the behavior of the field in these two sectors is responsible for
the (non)stabilization of the $L_2$ norms.

On the other hand, we found the right tendency of the behavior of the
solution for large $r$ what was the aim of the present calculations.

\section{Acknowledgment}

 The authors would like to thank Prof. V.B.Belyaev for the fruitful
 discussions. The work was partially supported by RFBR grant 08-01-00209.



\begin{thebibliography}{99}


\bibitem{Fad1} L. D. Faddeev, {\it Mathematical Aspects of the Three-Body Problem in the Quantum
Scattering Theory,} Trudy Matematicheskogo Instituta, v.69, (1963)
(in Russian), (Israel Program for Scientific Translations,
Jerusalem, (1965), (in English)).
\bibitem{enss} V.Enss, Ann.Phys., 119, 117-132, (1979).
\bibitem{derez} J.Derezinski and C.Gerard, J.Math.Phys.,
v.38(8), pp. 3925-3942, (1997).
\bibitem{BV} V.S.Buslaev,
A.F.Vakulenko, Unitary regularization for the three-body scattering,
Vestnik LGU, 13, 22--30, (1977).
\bibitem{ags67} E. O. Alt, P. Gra{\ss}berger, and W. Sandhas,
Nucl. Phys. {\bf B2}, 167, (1967).
\bibitem{p1} R.K. Peterkop, Zh. Eksp. Teor. Fiz, {\bf 43}, 616 (1962)
(in russian) [Sov. Phys. JETP {\bf 14}, 1377 (1962)].
\bibitem{p2} S.P. Merkuriev, Theor. Math. Phys., {\bf 32}, 680
(1977);
M. Brauner, J.S. Briggs and H.J. Klar,  J. Phys. B, {\bf 22}, 2265
(1989).
\bibitem{p3} E.O. Alt, A.M. Mukhamedzhanov,  JETP Lett.,
{\bf 56}, 435 (1992),  Phys. Rev. A, {\bf 47}, 2004 (1993); Y.E.
Kim, A.L. Zubarev, Phys. Rev. A, {\bf 56}, 521 (1997).
\bibitem{p4} J.H. Macek, S.Yu. Ovchinnikov,
 Phys. Rev. A, {\bf 54}, 1 (1996).
\bibitem{Rudge}  Rudge M R H 1968 {\it Rev. Mod. Phys.} {\bf 40}  564
\bibitem{Peterkop}  Peterkop R K  {\it Theory of Ionization of Atoms by Electron-Impact}
(Colorado Associated University Press, Boulder, 1977)
\bibitem{FaddeevMerkuriev}  Faddeev L D and  Merkuriev S P
{\it Quantum Scattering Theory for Several Particle Systems}
(Kluwer, Dordrecht, 1993)
\bibitem{Merkuriev} Merkuriev S P  1980 {\it Ann. Phys. (NY)} {\bf 130} 395
\bibitem{am92} E. O. Alt and A. M. Mukhamedzhanov, JETP Lett.
{\bf 56}, 435 (1992); Phys. Rev. A {\bf 47}, 2004 (1993).
\bibitem{asz78} E. O. Alt, W. Sandhas, and H. Ziegelmann,
Phys. Rev. C {\bf 17}, 1981 (1978).
\bibitem{AltLevinYakovlev}  Alt E O,  Levin S B and  Yakovlev S L
2004 {\it Phys. Rev.} C {\bf 69}  034002
\bibitem{as96} E. O. Alt and W. Sandhas, in {\it Coulomb Interactions in
Nuclear and Atomic Few-Body Collisions,} edited by F. S. Levin and
D. Micha (Plenum, New York 1996), p. 1.
\bibitem{Oryu} S.Oryu, S.Nishinohara, N.Shiiki, and S.Chiba,
Phys. Rev. C {\bf 75}, 021001(R) (2007).
\bibitem{Deltuva} A.Deltuva, A.C.Fonseca, P.U.Sauer,
Phys. Rev. C {\bf 71}, 054005 (2005).
\bibitem{kvr01} A. Kievsky, M. Viviani, and S. Rosati, Phys.
Rev. C {\bf 64}, 024002 (2001).
\bibitem{bly} V.B.Belyaev, S.B.Levin, S.L.Yakovlev,
    J.Phys. B, {\bf v.37}, 1369-1380, (2004)
\bibitem{Suslov} V.M.Suslov and B.Vlahovic,
   Phys.Rev. C, {\bf v.69}, 044003, (2004)
\bibitem{RescignoBray}  Rescigno T  N,  Baertschy M,  Isaacs W A and  McCurdy C W
1999 {\it Science} {\bf 286}  2474;  Baertschy M,  Rescigno T N and
McCurdy C W  2001 {\it Phys. Rev.} A {\bf 64}  022709;  Bray I
(2002) {\it Phys. Rev. Lett.} {\bf 89}  273201
\bibitem{BM} Buslaev, V. S.; Merkur'ev, S. P.,
Dokl. Akad. Nauk SSSR 189 269-272 (Russian); translated as Soviet
Physics Dokl. 14 (1969) 1055-1057
\bibitem{BMC}  Buslaev, V. S.; Merkuriev, S. P.; Salikov, S. P.
 Probl. Mat. Fiz., Leningrad. Univ.,
Leningrad, {\bf 9}, (1979), 14--30.
\bibitem{BMC1} Buslaev, V. S.; Merkuriev, S. P.; Salikov, S. P.  (Russian)  Zap.
Nauchn. Sem. Leningrad. Otdel. Mat. Inst. Steklov. (LOMI)  {\bf 84},
(1979), 16--22.
\bibitem{BL} V.S.Buslaev and S.B.Levin,
Amer.Math.Soc.Transl. (2)v.225, pp.55-71, (2008)
\bibitem{Veselova} A.M.Veselova, Theor.Math.Phys., 35(2), pp. 180--191, (1978).
\bibitem{Yang} C. N. Yang, Phys.
Rev. Lett. {\bf 19}, 1312 (1967).
\bibitem{Lieb} E. Lieb and W. Liniger, Phys.
Rev. {\bf 130}, 1605 (1963).
\bibitem{McGuire} J. B. McGuire, J.Math.Phys., {\bf 5}, 622 (1964).
\bibitem{Olshanii} M. Olshanii, Phys.Rev.Lett., {\bf 81},
 938 (1998).
\bibitem{Green} N. P. Mehta, B.D. Esry, and C.H.Green, Phys.Rev. A., {\bf 76},
 022711 (2007).
\bibitem{Gorl} A G\"orlitz et al., Phys.Rev.Lett., {\bf 87}, 130402 (2001).
\bibitem{K1} T. Kinoshita, T. Wenger and D. S. Weiss, Nature (London) {\bf 440},
900 (2006).
\bibitem{K2} T. Kinoshita, T. Wenger and D. S. Weiss, Science, {\bf 305}, 1125 (2004).
\bibitem{Esteve}  J. Esteve et al., Phys.Rev.Lett., {\bf 96}, 130403 (2006).

\end{thebibliography}
\end{document}